\documentclass[a4paper,11pt]{article}
\usepackage{amsmath}
\usepackage{amssymb}
\usepackage{amsfonts}
\usepackage[dvipdfmx]{graphicx}
\usepackage{color}
\usepackage{comment}
\usepackage{bm}
\usepackage[round]{natbib}

\newtheorem{definition}{Definition}
\newtheorem{theorem}{Theorem}
\newtheorem{lemma}[theorem]{Lemma}
\newtheorem{corollary}{Corollary}

\newtheorem{proposition}[theorem]{Proposition}

\title{\bf Priors for Reducing Asymptotic Bias of \\ the Posterior Mean}
\author{Yoichi Miyata \thanks{Faculty of Economics, Takasaki City University of Economics, 1300 Kaminamie, Takasaki, Gunma, 370-0801, Japan. E-mail: ymiyatagbt@tcue.ac.jp, ORCID:0000-0002-9822-690X}\and Takemi Yanagimoto \thanks{Institute of Statistical Mathematics, Tokyo, Japan}}

\begin{document}
\maketitle

\begin{abstract}
It is shown that the first-order term of the asymptotic bias of the posterior mean is removed by a suitable choice of a prior density. 
In regular statistical models including exponential families, and linear and logistic regression models, such a prior is given by the squared Jeffreys prior. 
We also explain the relationship between the proposed prior distribution, the moment matching prior, and the prior distribution for reducing the asymptotic bias of the posterior mode.
\end{abstract}
{\it Keywords}: Jeffreys prior, Posterior means, Prior elicitation, Unbiasedness;

\section{Introduction}\label{Intro}
Since 1990, Bayesian statistics has made great progress in both application and theory with the improvement of the computational power of computers. In Bayesian statistics, since parameters are treated as random variables, it is necessary to determine some prior distribution to obtain an estimator. 
Especially when the sample size is not so large, the Bayesian estimator is more likely to be affected by the choice of a prior distribution, and it becomes necessary to find a prior distribution that is valid in some sense. 
One of the prior distributions with such validity is the reference priors proposed by \cite{Be79} and \cite{BB89,BB92}. 
This prior distribution asymptotically maximizes the discrepancy between the prior and the posterior distribution as measured by Kullback-Leibler type divergence. This is based on the idea of choosing the least informative prior distribution to maximize the information obtained from the data.

On the other hand, another possible approach is to define the prior distribution so that the Bayesian estimator satisfies desirable properties in frequentist theory. 
\cite{GL11} proposed moment matching priors where the asymptotic error of order $n^{-1}$ between the obtained posterior mean and the maximum likelihood estimator (MLE) is zero. 
On the other hand, \cite{Fi93} studied a prior distribution such that the asymptotic bias (the first order bias) between the posterior mode and the true value is zero in the sense of frequency theory, and showed that under certain conditions, it is the prior distribution of \cite{Je46}. 

However, to the best of the author's knowledge, there has not been much research on the prior distribution such that the first-order bias between the posterior mean and the true value is zero. 
In this paper, we derive a prior distribution such that the asymptotic bias between the posterior mean and the true value is zero. We also show that the proposed prior is closely related to the bias reduction prior of \cite{Fi93} and the moment matching prior of \cite{GL11}.

In Section \ref{sec2}, we present a formal asymptotic expansion of the bias of the posterior mean for a true parameter vector and conditions for a prior distribution to remove the first order term in this bias. 
We also present some conditions for this prior to be the squared Jeffreys prior.
Section \ref{sec3} clarifies the relationship between the prior distribution derived in Section \ref{sec2}, the moment matching prior and the prior of \cite{Fi93}, and presents some implications for the main result of Section \ref{sec2} in terms of information geometry and equivalence of estimators.
In Section \ref{sec4}, we apply the proposed prior distributions for rather general families including the exponential distribution family and linear regression family, and in Section \ref{sec5}, for specific families of distributions such as the normal distribution and logistic regression model. 
Section \ref{sec6} gives some concluding remarks, and the Appendices contain proofs of theorems.

\section{Main results}\label{sec2}
In this section, we present a formal asymptotic expansion for the posterior mean bias. The rigorous conditions for the validity of the expansion are given in Appendices \ref{AppenA} and \ref{AppenB}.
Suppose that an observed random variable $Y_i$ has a probability density or mass function $p_{i}(y_{i}|\bm{\theta})$ where $\bm{\theta}$ is a parameter vector, and $Y_{1},\ldots ,Y_{n}$ are independent.
$\hat{\bm{\theta}}=(\hat{\theta}_{1},\ldots ,\hat{\theta}_{d})^{\top}$ denotes the maximum likelihood estimator (MLE) which maximizes the log-likelihood function $\ell (\bm{\theta}):=\sum_{i=1}^{n}\log p_{i}(y_{i}|\bm{\theta})$. 
The cumulants are defined as
\begin{align}
\kappa_{rs}&=\frac{1}{n}E_{\bm{\theta}}\left\{ \frac{\partial^{2}}{\partial \theta_{r}\partial_{s}}\ell(\bm{\theta})\right\}, \quad \kappa_{r,s}=\frac{1}{n}E_{\bm{\theta}}\left\{ \frac{\partial}{\partial \theta_{r}}\ell(\bm{\theta})\frac{\partial}{\partial \theta_{s}}\ell(\bm{\theta})\right\}, \notag \\
\kappa_{r,s,t}&=\frac{1}{n}E_{\bm{\theta}}\left\{ \frac{\partial}{\partial \theta_{r}}\ell(\bm{\theta})\frac{\partial}{\partial \theta_{s}}\ell(\bm{\theta})\frac{\partial}{\partial \theta_{t}}\ell(\bm{\theta})\right\} ,\quad \kappa_{rst}=\frac{1}{n}E_{\bm{\theta}}\left\{ \frac{\partial^3}{\partial \theta_{r}\partial \theta_{s}\partial \theta_{t}}\ell(\bm{\theta})\right\} , \notag \\
 &\textrm{and} \quad \kappa_{r,st}=\frac{1}{n}E_{\bm{\theta}}\left\{ \frac{\partial}{\partial \theta_{r}}\ell(\bm{\theta})\frac{\partial^2}{\partial \theta_{s}\partial \theta_{t}}\ell(\bm{\theta})\right\}, \label{cumulants}
\end{align}
where $E_{\bm{\theta}}\{ \cdot \}$ denotes the expectation under the density $p(\bm{y}|\bm{\theta})$ of $\bm{Y}=(Y_1,\ldots ,Y_n)^{\top}$. 
To make it explicit that these cumulants depend on the parameters, we sometimes write $\kappa_{r,s}=\kappa_{r,s}(\bm{\theta})$, $\kappa_{r,s,t}=\kappa_{r,s,t}(\bm{\theta})$, $\kappa_{r,st}=\kappa_{r,st}(\bm{\theta})$, and $\kappa_{rst}=\kappa_{rst}(\bm{\theta})$. In the cross-cumulant $\kappa_{r,st}$ when $Y_1,\ldots ,Y_n$ are independent, we have 
\begin{align*}
\kappa_{r,st}=\frac{1}{n}E_{\bm{\theta}}\left\{ \sum_{i=1}^{n}\frac{\partial}{\partial \theta_r}\log p_{i}(Y_i|\bm{\theta}) \frac{\partial^2}{\partial \theta_s \partial \theta_t}\log p_{i}(Y_i|\bm{\theta}) \right\} .
\end{align*}
Its proof is straightforward and will be omitted.

First, the asymptotic bias given by \cite{CS68} is rewritten in a cumulant-based form.
Let $\kappa^{r,s}$ be the $(r,s)$-element in the inverse of the matrix $(\kappa_{r,s})$. Then, the bias of the $k$th component $\hat{\theta}_k$ of the MLE is expressed as $E_{\bm{\theta}}\left\{ \hat{\theta}_k -\theta_k \right\} =B_{k,n}+O(n^{-2})$ where 
\begin{align}
B_{k,n}=\frac{1}{2n}\sum_{s,t,u}\kappa^{k,s}\kappa^{t,u}\left( \kappa_{stu}+2\kappa_{t,su}\right), \label{bias1}
\end{align}
and $B_{k,n}=O(n^{-1})$. The derivation of Equation \eqref{bias1} is given in Appendix \ref{asympt_bias_MLE}.

Next, we derive an asymptotic expansion for the posterior mean. 
When $\pi (\bm{\theta})$ is a prior density of a random parameter vector $\bm{\Theta}=(\Theta_1,\ldots ,\Theta_d )^{\top}$, the posterior mean of the $k$th component $\Theta_k$ has the form
\begin{align}
E_{\textrm{post}}[\bm{\Theta}_{k}]=\frac{\int \theta_{k}\exp\{ \ell(\bm{\theta})\} \pi (\bm{\theta})d\bm{\theta}}{\int \exp\{ \ell(\bm{\theta})\} \pi (\bm{\theta})d\bm{\theta}}. \label{posterior_mean}
\end{align}
For simplicity of exposition, we let $h(\bm{\theta})=-(1/n)\ell(\bm{\theta})$ and let the derivatives denoted by 
\begin{align*}
h_{r}(\bm{\theta})=\frac{\partial}{\partial \theta_{r}}h(\bm{\theta}), \quad 
h_{rs}(\bm{\theta})=\frac{\partial^2}{\partial \theta_{r}\partial\theta_s}h(\bm{\theta}),\quad \textrm{and}\quad h_{rsj}(\bm{\theta})=\frac{\partial^3}{\partial \theta_{r}\partial\theta_s \partial \theta_j}h(\bm{\theta}).
\end{align*}
Similarly, we let $\pi_{j}(\bm{\theta})=(\partial /\partial \theta_{j})\pi(\bm{\theta})$ and $g_{ij}(\bm{\theta})=(\partial^{2}/\partial \theta_{i}\partial\theta_{j})g(\bm{\theta})$. 
The value of function $f(\bm{\theta})$ at $\bm{\theta}=\hat{\bm{\theta}}$ is abbreviated as $\hat{f}$. For example, $\hat{h}_{ij}=h_{ij}(\hat{\bm{\theta}})$ and $\hat{\pi}=\pi (\hat{\bm{\theta}})$. Let $\hat{h}^{ij}$ be the $(i,j)$-element of the inverse matrix of the Hessian $(\hat{h}_{ij})$.

From the standard-form Laplace approximation (2.6) of \cite{KTK90}, the posterior mean \eqref{posterior_mean} has an expansion
\begin{align}
E_{\textrm{post}}\left[ \Theta_{k}\right] &=\hat{\theta}_{k}+\frac{1}{n}\sum_{j}\hat{h}^{kj}\left\{ \frac{\hat{\pi}_{j}}{\hat{\pi}} -\frac{1}{2}\sum_{r,s}\hat{h}^{rs}\hat{h}_{rsj}\right\} +\frac{R_{1n}}{n^{2}}, \label{Laplace1}
\end{align}
where $R_{1n}$ indicates an asymptotic error term which is allowed to depend on the sample $\bm{Y}$, and $R_{1n}=O_{p}(1)$ holds under the conditions in Appendix \ref{AppenA}.

Equation \eqref{Laplace1} can be expressed as
\begin{align}
E_{\textrm{post}}\left[ \Theta_{k}\right] &=\hat{\theta}_{k}+\frac{1}{n}\sum_{j}\kappa^{k,j}\left\{ \frac{\partial}{\partial \theta_{j}}\log \pi(\bm{\theta}) +\frac{1}{2}\sum_{r,s}\kappa^{r,s}\kappa_{rsj}\right\} +\frac{R_{2n}}{n}+\frac{R_{1n}}{n^{2}}, \label{Laplace2}
\end{align}
where 
\begin{align}
R_{2n}=\sum_{j}\hat{h}^{kj}\left\{ \frac{\hat{\pi}_{j}}{\hat{\pi}} -\frac{1}{2}\sum_{r,s}\hat{h}^{rs}\hat{h}_{rsj}\right\}-\sum_{j}\hat{h}^{kj}\left\{ \frac{\hat{\pi}_{j}}{\hat{\pi}} -\frac{1}{2}\sum_{r,s}\kappa^{r,s}\kappa_{rsj}\right\} .
\label{R2n}
\end{align}
Note that under a suitable assumption, it holds that $R_{2n}=o_{p}(1)$. 

We now expand the bias of the posterior mean \eqref{posterior_mean}. 
Subtracting $k$th component $\theta_{k}$ of the true parameter vector from both the sides of Equation \eqref{Laplace2} and using the first-order bias \eqref{bias1} yields
\begin{align}
E_{\textrm{post}}\left[ \Theta_{k}\right] -\theta_{k}&=\hat{\theta}_{k}-\theta_{k}-B_{k,n} \notag\\
 &\quad +\frac{1}{n}\sum_{j}\kappa^{k,j}\left\{ \frac{\partial}{\partial \theta_{j}}\log \pi (\bm{\theta}) +\frac{1}{2}\sum_{r,s}\kappa^{r,s}\kappa_{rsj}\right\} +B_{k,n}+\frac{R_{2n}}{n}+\frac{R_{1n}}{n^{2}} \notag\\
 &=\hat{\theta}_{k}-\theta_{k}-B_{k,n} \notag\\
 &\quad +\frac{1}{n}\sum_{j}\kappa^{k,j}\left\{ \frac{\partial}{\partial \theta_{j}}\log \pi (\bm{\theta}) +\frac{1}{2}\sum_{r,s}\kappa^{r,s}(\kappa_{rsj}+\kappa_{jrs}+2\kappa_{r,js})\right\} +\frac{R_{2n}}{n}+\frac{R_{1n}}{n^{2}} \notag \\
&=\hat{\theta}_{k}-\theta_{k}-B_{k,n} \notag\\
 &\quad +\frac{1}{n}\sum_{j}\kappa^{k,j}\left\{ \frac{\partial}{\partial \theta_{j}}\log \pi (\bm{\theta}) +\sum_{r,s}\kappa^{r,s}(\kappa_{rsj}+\kappa_{r,js})\right\} +\frac{R_{2n}}{n}+\frac{R_{1n}}{n^{2}}. \notag 
\end{align}
Consequently, we have the following theorem.
\begin{proposition}\label{asympt.expans}
Under conditions [A1]--[A8], the bias of the posterior mean \eqref{posterior_mean} is expressed as
\begin{align}
E_{\bm{\theta}}\left\{ E_{\textrm{post}}[\Theta_k]-\theta_k \right\} &=\frac{1}{n}\sum_{j}\kappa^{k,j}\left\{ \frac{\partial}{\partial \theta_{j}}\log \pi (\bm{\theta}) +\sum_{r,s}\kappa^{r,s}(\kappa_{rsj}+\kappa_{r,js})\right\} +o\left( \frac{1}{n}\right) . \label{main_result}
\end{align}
\end{proposition}
Conditions [A1]--[A8] are given in Appendices \ref{AppenA} and \ref{AppenB}. Here we consider a prior distribution that eliminates the term of order $n^{-1}$.
\begin{definition}\label{def1}
We say that $\pi (\bm{\theta})$ is a bias reduction prior if $\pi (\bm{\theta})$ satisfies for any $j=1,\ldots ,d$
\begin{align}
\frac{\partial}{\partial \theta_{j}}\log \pi (\bm{\theta}) +\sum_{r,s}\kappa^{r,s}(\bm{\theta})(\kappa_{rsj}(\bm{\theta})+\kappa_{r,js}(\bm{\theta}))=0 . \label{def_eq1}
\end{align}
\end{definition}
If Equation \eqref{def_eq1} holds, then $E_{\bm{\theta}}\{ E_{\textrm{post}}[\Theta_k]-\theta_k \} =o\left( n^{-1}\right)$. Surprisingly, Equation \eqref{def_eq1} does not depend on the subscript $k$ indicating the component of the parameter vector. Hereafter, the prior distribution satisfying Equation \eqref{def_eq1} will be denoted by $\pi_{BR}(\bm{\theta})$.
It is not clear in general whether there exists a prior distribution satisfying Equation \eqref{def_eq1}. However, it always exists and can be expressed explicitly when the parameters are one-dimensional.
\begin{corollary}\label{coro1}
Assume that $\Theta$ is one dimensional, $Y_{1},\ldots ,Y_{n}$ are independently and identically distributed (i.i.d.) with density $p(y|\theta)$, and the interchange of integral and derivative is permissible, e.g., $(d/d\theta )\int \{(d^{2}/d\theta^{2})\log p(y|\theta)\} p(y|\theta )dy=\int (d /d\theta ) \left[\left\{ (d^{2}/d\theta^{2})\log p(y|\theta)\right\} p(y|\theta ) \right] dy$. 
Then, the bias reduction prior satisfying Equation \eqref{def_eq1} is proportional to the Fisher information $I_{1}(\theta )=-E_{\theta}\left\{ (\partial^2/\partial \theta^2 )\log p(Y_{1}|\theta)\right\}$, that is $\pi_{BR} (\theta)\propto I_{1}(\theta )$.
\end{corollary}

Several families of probability distributions including the exponential distribution family satisfy the following condition:
\begin{description}
\item[(C)] For any $r, s, t \in \{ 1,\ldots ,d\}$, $\kappa_{r,st}=0$.
\end{description}
In this case, the bias-reduction prior given in Definition \ref{def1} can be expressed in a simple form as below.
\begin{corollary}\label{coro2}
Under condition (C), the bias reduction prior satisfying Equation \eqref{def_eq1} is given by 
\begin{align*}
\pi_{BR} (\bm{\theta})\propto |\bm{I}(\bm{\theta})|,
\end{align*}
where $|\cdot |$ indicates the determinant of a matrix, and $\bm{I}(\bm{\theta})=(-\kappa_{rs})$ is the Fisher information matrix based on the density or mass function $p(\bm{y}|\bm{\theta})$ of $\bm{Y}=(Y_1,\ldots ,Y_n)^{\top}$.
\end{corollary}
Although the Fisher information matrix $\bm{I}(\bm{\theta})$ may depend on the sample size $n$, the subscript $n$ is omitted here.
To yield another corollary, we rewrite $\kappa_{rsj}(\bm{\theta})+\kappa_{r,js}(\bm{\theta})$ in the second term of Equation \eqref{def_eq1} as a single-term expression 
\begin{align}
\frac{\partial}{\partial \theta_{r}}\kappa_{js}(\bm{\theta}) \label{kappa_bibun}
\end{align}
This expression leads us to the following corollary.
\begin{corollary}\label{coro3}
A sufficient condition for the second term of Equation \eqref{def_eq1} to vanish is that the Fisher information matrix $\bm{I}(\bm{\theta})$ is independent of $\bm{\theta}$.
\end{corollary}
When the Fisher information matrix is independent of $\bm{\theta}$, $\pi_{BR} (\bm{\theta})$ is the uniform prior, that is, $\pi_{BR} (\bm{\theta})\propto 1$. It is also the squared Jeffreys prior.

\section{Implications}\label{sec3}
Possible implications of the main results are discussed here. They consist of three notable points.
\subsection{Two related priors}
To aid our better understanding of $\pi_{BR}(\bm{\theta})$, we examine the existing two priors by \cite{Fi93} and \cite{GL11}. 
In this section, for the sake of brevity, we assume that the samples $Y_1,\ldots ,Y_n$ are i.i.d. as in \cite{GL11}. 
The bias reduction prior of the posterior mode, $\pi_{BM}(\bm{\theta})$, was introduced in \cite{Fi93}, where the prior was treated as a penalized likelihood in the frequentist framework. The equation in the 17th line from the bottom on page 29 of \cite{Fi93} is written in the present
notation as
\begin{align}
\frac{\partial}{\partial \theta_{j}}\log \pi_{BM}(\bm{\theta})-\frac{1}{2}\sum_{r=1}^{d}\sum_{s=1}^{d}\kappa^{r,s}\left(\kappa_{j,r,s}+\kappa_{j,rs}\right) =0 \quad (j=1,\ldots ,d). \label{firth_prior0}
\end{align}
He emphasized the equivalency relationship between $\pi_{BM}(\bm{\theta})$ and the Jeffreys prior when the sampling density is in the exponential family. Applying the Bartlett identity $\kappa_{jrs}+\kappa_{j,rs}+\kappa_{r,js}+\kappa_{s,jr}+\kappa_{j,r,s}=0$ to this equation, we obtain another form of $\pi_{BM}(\bm{\theta})$, 
\begin{align}
\frac{\partial}{\partial \theta_{j}}\log \pi_{BM}(\bm{\theta})+\sum_{r=1}^{d}\sum_{s=1}^{d}\kappa^{r,s}\left( \frac{1}{2}\kappa_{jrs}+\kappa_{r,js}\right) =0\quad (j=1,\ldots ,d). \label{firth_prior}
\end{align}
On the other hand, from the equation in the third line from the bottom on page 193 of \cite{GL11}, the moment matching prior fulfills that for any $j = 1,\ldots ,d$, 
\begin{align}
\frac{\partial}{\partial \theta_{j}}\log \pi_{MM}(\bm{\theta})+\frac{1}{2}\sum_{r=1}^{d}\sum_{s=1}^{d}\kappa_{jrs}\kappa^{r,s}=0. \label{MM_prior}
\end{align}
From Equations \eqref{firth_prior}, \eqref{MM_prior} and \eqref{def_eq1}, we obtain the following relationship.
\begin{proposition}\label{prop1}
Supposed that the samples $Y_1,\ldots ,Y_n$ are i.i.d. and there exist priors $\pi_{BR}$ and $\pi_{BM}$ satisfying Equations \eqref{def_eq1} and \eqref{firth_prior} respectively. Then, it holds that $\pi_{BR}(\bm{\theta})= \pi_{BM}(\bm{\theta})\pi_{MM}(\bm{\theta})$ for every $\bm{\theta}$.
\end{proposition}
The proof is obvious from Equations \eqref{firth_prior}, \eqref{MM_prior} and \eqref{def_eq1}, and is therefore omitted. 
We also consider the expression of the bias reduction prior using the coefficients given in the information geometry. 
The statistical cube tensor and the e-connection coefficients are defined as 
\begin{align*}
T_{rsj}&=E_{\bm{\theta}}\left[\frac{\partial}{\partial \theta_{r}}\ell (\bm{\theta})\frac{\partial}{\partial \theta_{s}}\ell (\bm{\theta})\frac{\partial}{\partial \theta_{j}}\ell (\bm{\theta})\right] , \text{ and } \overset{(e)}{\Gamma}_{rs,j}=E_{\bm{\theta}}\left[ \frac{\partial^2}{\partial\theta_{r}\partial\theta_{s}}\ell (\bm{\theta})\frac {\partial}{\partial\theta_{j}}\ell (\bm{\theta})\right]
\end{align*}
In addition, the $\alpha$-connection coefficients are defined as
\begin{align*}
\overset{(\alpha )}{\Gamma}_{rs,j}=\overset{(e)}{\Gamma}_{rs,j}+\frac{1-\alpha}{2}T_{rsj}.
\end{align*}
When $\alpha =-1$, $\overset{(-1)}{\Gamma}_{rs,j}$ is called $m$-connection coefficient. Hereafter, the m-connection coefficient will be denoted by the symbol $\overset{(m)}{\Gamma}_{rs,j}$, which is a commonly used notation. To save space here, we omit the geometrical interpretation. See, for example, \cite{AN07}. 

In this case, \cite{Ta23} shows in terms of the connection coefficient that the moment matching prior of \cite{GL11} fulfills
\begin{align*}
\frac{\partial}{\partial \theta_{j}}\log \pi_{MM}(\bm{\theta})
-\left\{ \frac{\partial}{\partial \theta_{j}}\log \pi_{J}(\bm{\theta})+\frac{1}{2}\sum_{r=1}^{d}\sum_{s=1}^{d}\kappa^{r,s}\overset{(e)}{\Gamma}_{rs,j}\right\} =0,
\end{align*}
where $\pi_{J}(\bm{\theta})$ is the Jeffreys prior.
On the other hand, from equation \eqref{firth_prior0}, the prior density of \cite{Fi93} satisfies $(\partial /\partial \theta_{j})\log \pi(\bm{\theta})-(1/2)\sum\sum_{r,s}\kappa^{r,s}\overset{(m)}{\Gamma}_{rs,j}=0$. 
From these two equations, the bias reduction prior satisfies
\begin{align*}
\frac{\partial}{\partial \theta_{j}}\log \pi_{BR}(\bm{\theta})
-\left\{ \frac{\partial}{\partial \theta_{j}}\log \pi_{J}(\bm{\theta})+\frac{1}{2}\sum_{r=1}^{d}\sum_{s=1}^{d}\kappa^{r,s}(\overset{(e)}{\Gamma}_{rs,j}+\overset{(m)}{\Gamma}_{rs,j})\right\} =0 .
\end{align*}

\subsection{Asymptotical equivalence order among Bayesian estimators}
Notable asymptotical equivalencies hold among Bayesian estimators induced from the priors in the study. 
The moment matching prior $\pi_{MM}(\bm{\theta})$ was originally designed for pursuing a noninformative prior under which the posterior mean $\hat{\bm{\theta}}_{MM}$ is asymptotically equivalent with the MLE $\hat{\bm{\theta}}_{ML}$. Their interest focused on the case where the asymptotical order $-3/2$, that is
\begin{align}
\| \hat{\bm{\theta}}_{MM} - \hat{\bm{\theta}}_{ML} \| = O_{p}(n^{-3/2}),
\end{align}
where the symbol $\| \cdot \|$ stands for the Euclidean norm. 
A higher order asymptotic equivalence of $O_{p}(n^{-2})$ is observed in several familiar models. An example is the case of the exponential family with the canonical parameter $\bm{\theta}$ \cite[]{YO21}. More generally, we write the order of the asymptotic equivalency between them as $O_{p}(n^{-\alpha})$.
A general sufficient condition on the asymptotic equivalence between the posterior mean and the posterior mode was given by \citet[Theorem 2.1]{YM24}. Consider three prior functions, $\pi_{A}(\bm{\theta})$, $\pi_{r}(\bm{\theta})$ and $\pi_{N}(\bm{\theta})$ satisfying the equality $\pi_{A}(\bm{\theta}) = \pi_{r}(\bm{\theta})\pi_{N}(\bm{\theta})$ holds for every $\bm{\theta}$. They showed, under the weak regularity conditions the asymptotic equality
\begin{align*}
\| (\hat{\bm{\theta}}_{A} - \hat{\bm{\theta}}_{N} ) - (\hat{\bm{\theta}}_{r} - \hat{\bm{\theta}}_{ML}) \| = O_{p}(n^{-2}),
\end{align*}
where $\hat{\bm{\theta}}_{A}$ and $\hat{\bm{\theta}}_{N}$ are the posterior means under the priors $\pi_{A}(\bm{\theta})$ and $\pi_{N}(\bm{\theta})$, and $\pi_{r}(\bm{\theta})$ is the posterior mode under the prior $\hat{\bm{\theta}}_{r}$. 
To apply their result, we set $\pi_{A}(\bm{\theta})$, $\pi_{r}(\bm{\theta})$ and $\pi_{N}(\bm{\theta})$ as $\pi_{BR}(\bm{\theta})$, $\pi_{BM}(\bm{\theta})$ and $\pi_{MM}(\bm{\theta})$, respectively. We examine the asymptotic relationship among the four estimators; the posterior means $\hat{\bm{\theta}}_{BR}$ under $\pi_{BR}(\bm{\theta})$, the posterior mode $\hat{\bm{\theta}}_{BM}$ under $\pi_{BM}(\bm{\theta})$, the posterior mean $\hat{\bm{\theta}}_{MM}$ under $\pi_{MM}(\bm{\theta})$ and the MLE $\hat{\bm{\theta}}_{ML}$. It follows for $3/2\leq \alpha \leq 2$ that $\|\hat{\bm{\theta}}_{BR} - \hat{\bm{\theta}}_{BM}\| = O(n^{-\alpha})$ if $\|\hat{\bm{\theta}}_{MM} - \hat{\bm{\theta}}_{ML}\|= O(n^{-\alpha})$. 
We can expect an asymptotic equivalency order between $\hat{\bm{\theta}}_{BR}$ and $\hat{\bm{\theta}}_{BM}$ is high, though the order depends on the family of sampling densities. 
An implication of the present view pertains to the dependence between the choice between two priors and that between the posterior mean and the posterior mode. A pair of choices are required to seek the asymptotically equivalent estimators.
\subsection{Role of bias reduction}
Both the priors, $\pi_{BR}(\bm{\theta})$ and $\pi_{BM}(\bm{\theta})$, are designed for eliminating the first order asymptotic biases of the posterior mean $\hat{\bm{\theta}}_{BR}$ under the former prior and the posterior mode $\hat{\bm{\theta}}_{BM}$ under the latter prior, respectively. Recall that the primary aim of constructing a prior in \cite{Fi93} was to yield a second order asymptotically efficient estimator in the frequentist context. 
Here we note the difference between the optimality properties of the posterior mean and the posterior mode. The former is the minimization of the posterior mean of the quadratic loss $E_{\textrm{post}}[\| \hat{\theta} -\Theta \|^2]$, and the latter is that of the zero-one loss. The former loss is closely related with the mean squared error, which is decomposed into the squared bias and the variance. This view indicates that the amount of bias of an estimator becomes critical when the quadratic loss is regarded as a desired one.
A detailed case-by-case comparative study of $\hat{\bm{\theta}}_{BR}$ and $\hat{\bm{\theta}}_{BM}$ would be needed. Regarding the comparative analysis presented in the following examples, the former would be more promising.

\section{Examples of general families}\label{sec4}
\subsection{Multivariate location families}\label{example4.1}
Suppose that $d$-dimensional observed random vectors $\bm{Y}_1,\dots ,\bm{Y}_n$ are i.i.d. with a density $p(\bm{y}|\bm{\mu})=g(\bm{y}-\bm{\mu})$ where $\bm{\mu}=(\mu_1,\ldots ,\mu_{d})^{\top}$ is an unknown parameter vector and $g:\mathbb{R}^{d}\to [0,\infty)$ is a smooth function.
We write the $(j,k)$-element in the Fisher information matrix for the density $p(\bm{y}|\bm{\mu})$ as $I_{jk}$, and assume that the matrix $(I_{jk})$ is nonsingular. By using the chain rule, 
\begin{align*}
I_{jk}&=\int \frac{\partial}{\partial \mu_{j}}\log g(\bm{y}-\bm{\mu})\left\{ \frac{\partial}{\partial \mu_{k}}\log g(\bm{y}-\bm{\mu})\right\} g(\bm{y}-\bm{\mu})d\bm{y} \\
  &=\int_{\mathbb{R}^{d}} \frac{\partial}{\partial z_{j}}\log g(\bm{z})\left\{ \frac{\partial}{\partial z_{k}}\log g(\bm{z})\right\} g(\bm{z})d\bm{z},
\end{align*}
which is independent of the parameter vector $\bm{\mu}$.
By Corollary \ref{coro3}, the bias reduction prior is given by $\pi_{BR}(\bm{\mu})\propto 1$.

\subsection{Linear regression model with the location parameter}\label{example4.2}
The location parameter with the parameter space $(-\infty ,\infty)$ provides us with a tractable linear regression model. Note that generalized linear regression models often have serious problems due to the restricted parameter space. 
We consider a simple and powerful linear regression model with a $p$-dimensional parameter vector $\bm{\beta}$. Let $Y_i$ be a response variable and let $\bm{z}_i^{\top}$ be the $i$-th row vector of the design matrix $\bm{Z}$, where $\bm{Z}^{\top}\bm{Z}$ is assumed to be non-singular. A convenient form for the density of $Y_i$ is 
\begin{align*}
p(y_{i}|\bm{\beta}) = \exp \left\{ g(y_i - \bm{z}_i^{\top}\bm{\beta})\right\} ,\quad  (i = 1,\ldots ,n)
\end{align*}
where $g(x)$ is a smooth real-valued function. When letting $g^{\prime\prime}(x)=(d^{2}/dx^{2})g(x)$, the Fisher information matrix is expressed as $\bm{I}(\bm{\beta}) = -c\bm{Z}^{\top}\bm{Z}$, with $c = \int g^{\prime\prime}(x) \exp(g(x))dx$, which is independent of $\bm{\beta}$. 
It follows from Corollary \ref{coro3} that $\pi_{BR}(\bm{\beta})$ is uniform  for $\bm{\beta}$, which is also the Jeffreys prior. Amazingly, this prior elicitation is free from the choice of $g(x)$ in the regression model.

\subsection{Exponential families}
Suppose that observed random variables $Y_1,\dots ,Y_n$ are i.i.d. with a density in the canonical form,
\begin{align}
p(y|\bm{\theta})&=a(y)c(\bm{\theta})\exp\left\{ \sum_{j=1}^{k}\theta_{j}T_{j}(y)\right\}
\end{align}
with respect to a $\sigma$-finite measure, where $a(y)$ and $c(\bm{\theta})$ are real-valued functions of $y$ and $\bm{\theta}$ and $\bm{\theta}=(\theta_1,\ldots ,\theta_k )^{\top}$ is an unknown parameter vector.
Then, the density of $\bm{Y}=(Y_1,\ldots ,Y_n)^{\top}$ is written as
\begin{align*}
p(\bm{y}|\bm{\theta})=c(\bm{\theta})^n \exp\left\{ \sum_{j=1}^{k}\theta_{j}S_{j}(\bm{y})\right\} \prod_{i=1}^{n}a(y_{i}),
\end{align*}
where $S_{j}(\bm{y})=\sum_{i=1}^{n}T_{j}(y_{i})$.
As the log-likelihood function becomes 
\begin{align*}
\ell (\bm{\theta})&=n\log c(\bm{\theta})+\sum_{j=1}^{k}\theta_{j}S_{j}(\bm{y})+\sum_{i=1}^{n}\log a(y_{i}),
\end{align*}
the Hessian of minus the log-likelihood function is given by 
\begin{align*}
-\frac{\partial^2}{\partial \bm{\theta}\partial \bm{\theta}^{\top}}\ell(\bm{\theta})&=-n\frac{\partial^2}{\partial \bm{\theta}\partial \bm{\theta}^{\top}}\log c(\bm{\theta}).
\end{align*}
Because this does not include any random variables, it satisfies the assumption (C). 
Accordingly, by Corollary \ref{coro2}, the bias-reduction prior is given by $\pi_{BR} (\bm{\theta})\propto \left|-(\partial^2 /\partial \bm{\theta}\partial \bm{\theta}^{\top})\log c(\bm{\theta})\right|$. 
Note that this squared Jeffreys prior is equivalent to the uniform prior for the expectation parameter $E_{\bm{\theta}}\{ (T_{1}(Y_{1}),\ldots ,T_{k}(Y_{1}))^{\top}\}$.
\section{Examples of specific families}\label{sec5}

\subsection{Normal distribution}
Consider that observed random variables $Y_1,\ldots ,Y_n$ are i.i.d. according to Normal distribution $N(\mu,\xi)$ with mean $\mu$ and variance $\xi$. Set $\bm{\theta}= (\mu,\xi)^{\top}$. $\pi_{BR}(\bm{\theta})$ is a solution to the partial differential equation \eqref{def_eq1}, that is $\partial \log \pi (\bm{\theta})/\partial\mu = 0$ and $\partial \log \pi (\bm{\theta})/\partial \xi = -2/\xi$, which yields that $\pi_{BR}(\bm{\theta})\propto \xi^{-2}$. The resultant posterior mean is expressed as $(\hat{\mu}, \hat{\xi}) = (\bar{Y}, s^2)$ with $\bar{Y}=n^{-1}\sum_{i=1}^{n}Y_{i}$ and $s^2=(n-1)^{-1}\sum_{i=1}^{n}(Y_{i}-\bar{Y})^2$ being the unbiased estimator of $\xi$.
 The need for bias reduction becomes evident, when the population distribution is the multiple normal distribution with $K$ strata; for each $k = 1,\ldots ,K$, $Y_{k1},\ldots ,Y_{k\, n_{k}}$ are i.i.d. with $N(\mu_k,\xi )$. Set $\bm{\theta}= (\mu_1,\ldots ,\mu_K, \xi)^{\top}$. Routine calculations yield that $\pi_{BR}(\bm{\theta})\propto \xi^{-2}$, which is independent of $K$. Similarly, it follows that $\pi_{BM}(\bm{\theta})\propto \xi^{K/2}$, which depends on $K$. Both the induced estimators of $\xi$, $\hat{\xi}_{BR}$ and $\hat{\xi}_{BM}$ are commonly equal to $s_{G}^{2}=\sum_{k}\sum_{i} (Y_{ki} -\bar{Y}_{k})^{2}/(N - k)$ with $N =\sum_{k} n_{k}$ and $\bar{Y}_{k}=(1/n_{k})\sum_{i=1}^{n_{k}}Y_{ki}$. It looks that the prior $\pi_{BM}(\bm{\theta})$ places unreasonably heavy weights on large values of $\xi$.
Recall that the reference prior, a familiar noninformative prior in this model, is proportional to $\xi^{-1}$ and independent of $K$. 
The posterior mean of the canonical parameter $\bm{\theta}= (\mu_{1}/\xi,\ldots,\mu_{K}/\xi,1/\xi)$ results in the equivalent estimator of $\xi$ with $s_{G}^2$. When $K$ is large, we observe that $\pi_{BM}(\bm{\theta})$ is isolated from the other two priors.

\subsection{Logistic regression model}
Finally, we consider the logistic regression model. 
Each of dependent variables $Y_i$ $(i=1,\ldots , n)$ has a probability mass function 
\begin{align*}
p_{i}(y_i|\bm{\beta})&=F(\bm{x}_{i}^{\top}\bm{\beta})^{y_i}\left( 1-F(\bm{x}_{i}^{\top}\bm{\beta}) \right)^{1-y_i}, \quad (y_i \in\{ 0,1 \}),
\end{align*}
where $F(t)=\exp(t)/(1+\exp(t))$, $\bm{x}_i =(x_{i1},\ldots ,x_{id})^{\top}$ is a covariate vector, and $\bm{\beta}=(\beta_1\ldots ,\beta_{d})^{\top}$ is an unknown coefficient vector. 
As the log-likelihood function is 
\begin{align*}
\ell (\bm{\beta})&=\sum_{i=1}^{n}\left[ y_i\bm{x}_{i}^{\top}\bm{\beta}-\log\left\{ 1+\exp (\bm{x}_{i}^{\top}\bm{\beta})\right\}\right],
\end{align*}
the Hessian is given by 
\begin{align}
\frac{\partial^2}{\partial \bm{\beta}\partial \bm{\beta}^{\top}}\ell (\bm{\beta})&=-\bm{X}^{\top}W(\bm{\beta}) \bm{X}, \label{Hessian1}
\end{align}
where $\bm{X}^{\top}=(\bm{x}_1,\ldots ,\bm{x}_n )$ is a $d\times n$ covariate matrix, $F^{\prime}(t)=(d/dt)F(t)=F(t)(1-F(t))$, $\textrm{diag}\{ \cdot \}$ stands for the diagonal matrix, and $W(\bm{\beta})=\textrm{diag}\{ F^{\prime}(\bm{x}_{1}^{\top}\bm{\beta}),\ldots,F^{\prime}(\bm{x}_{n}^{\top}\bm{\beta}) \}$. 
Because the Hessian \eqref{Hessian1} does not depend on any random variable, this model satisfies the assumption (C).
Accordingly, by Corollary \ref{coro2}, the bias-reduction prior is given by $\pi_{BR} (\bm{\beta})\propto |\bm{X}^{\top}W(\bm{\beta})\bm{X}|$. 
To examine how much the proposed prior distribution improves the bias of the posterior means of the parameters, we consider the following logistic regression model with $d=3$. 
\begin{align}
\textrm{logit}(F(\bm{x}_{i}^{\top}\bm{\beta}))=-1.25x_{i1}+0.75x_{i2}+0.2x_{i3}\quad (i=1,\ldots ,n).
\end{align}
Note that $\bm{\beta}_{0}=(-1.25,0.75,0.2)^{\top}$ is a true value parameter vector.
The explanatory variables $x_{1i},x_{2i},x_{3i}$ $(i=1,\ldots ,30)$ are generated from the trivariate normal distribution $N_{3}(\bm{0}_{3},\bm{\Sigma})$ where $\bm{0}_{3}=(0,0,0)^{\top}$, $\rho=0.1$, and 
\begin{align*}
\bm{\Sigma}=\begin{pmatrix} 
 1 & \rho & \rho^{2} \\
\rho & 1 & \rho\\
\rho^{2} & \rho & 1 
\end{pmatrix}
\end{align*}
We generated binary random numbers $y_{i}$ $(i=1,\ldots ,30)$ from the Bernoulli distribution with a probability of success $F(\bm{x}_{i}^{\top}\bm{\beta})$. 
Here, we compare the following three prior distributions
\begin{align}
\pi_{BR}(\bm{\beta})\propto\left| \bm{X}^{\top}W(\bm{\beta})\bm{X}\right| ,\quad \pi_{BM}(\bm{\beta})\propto\left| \bm{X}^{\top}W(\bm{\beta})\bm{X}\right|^{1/2} , \quad \textrm{and } \pi_{U}(\bm{\beta})\propto 1. \label{prior3}
\end{align}
The second one is the Jeffreys prior which is equivalent to that of \cite{Fi93} and the third one is the uniform prior. 
The posterior density function of $\bm{\beta}$ was derived based on each prior distribution, and the Markov chain Monte Carlo (MCMC) method was used to generate random numbers for $\bm{\beta}$. 
To implement the MCMC method, we applied the Metropolis-within-Gibbs method \citep{Mu91}, in which a candidate sample is generated by a random walk chain, and the maximum likelihood estimator is set as an initial value in each parameter. 
The generated MCMC samples of size 10000 were used to compute the posterior means and the biases. 

\begin{figure}[htb]
\begin{minipage}{.33\linewidth}
\includegraphics[width=\linewidth]{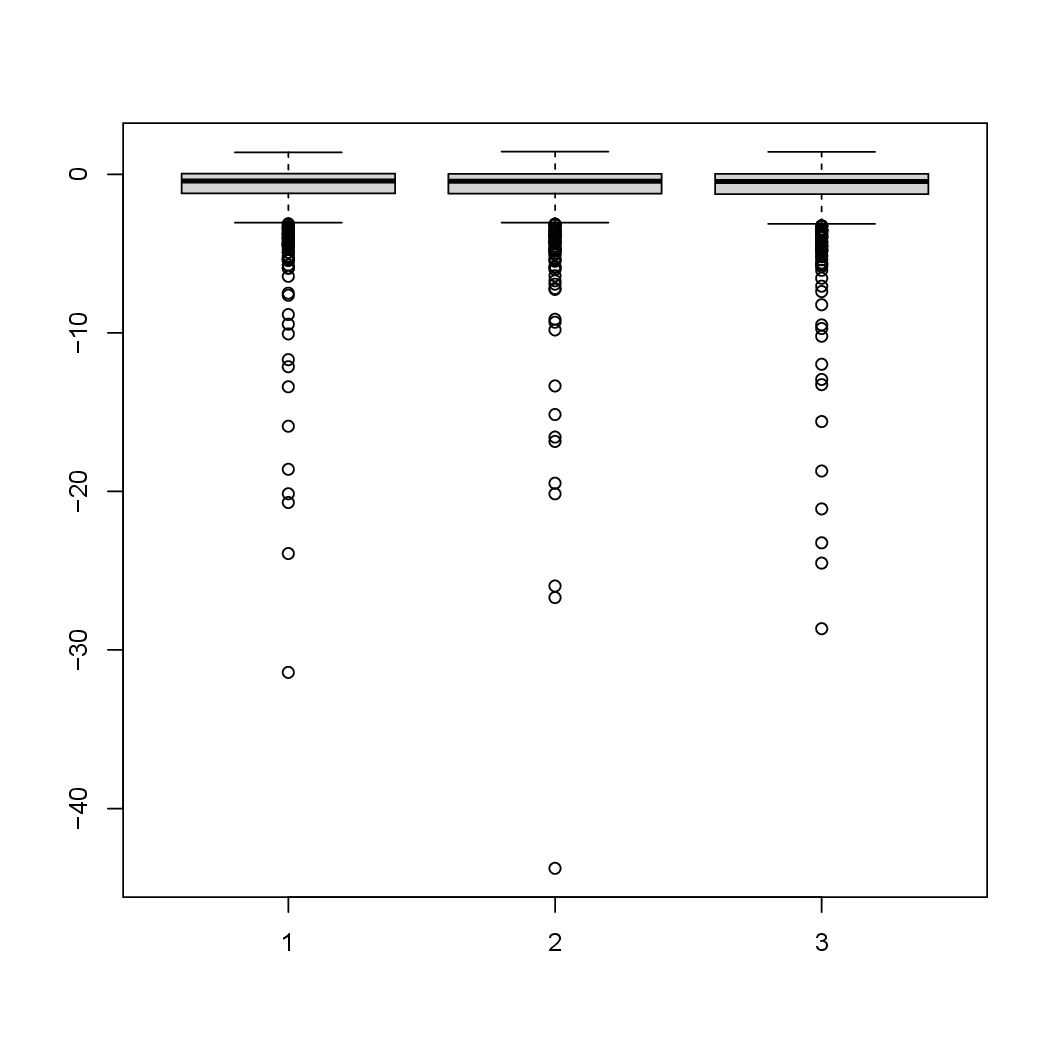}
\end{minipage}
\begin{minipage}{.33\linewidth}
\includegraphics[width=\linewidth]{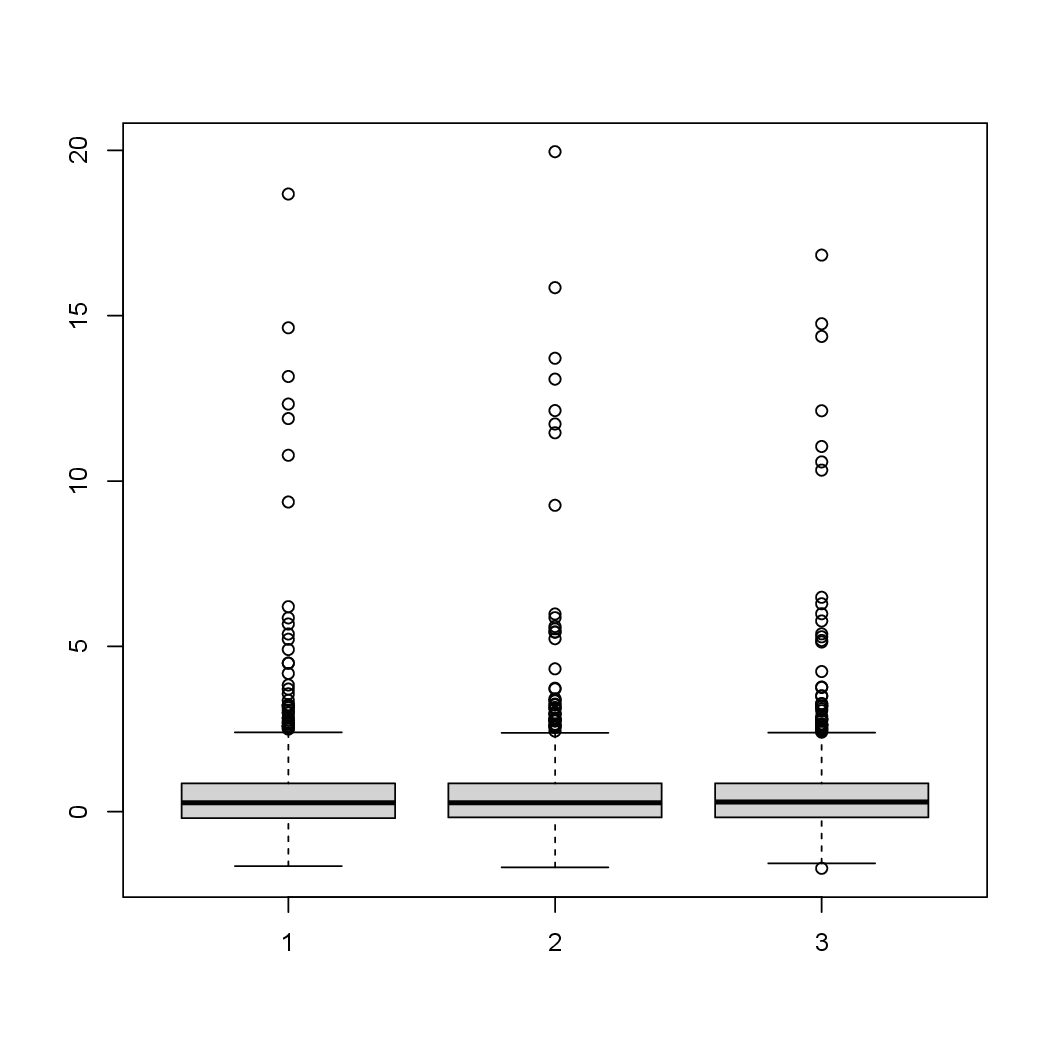}
\end{minipage}
\begin{minipage}{.33\linewidth}
\includegraphics[width=\linewidth]{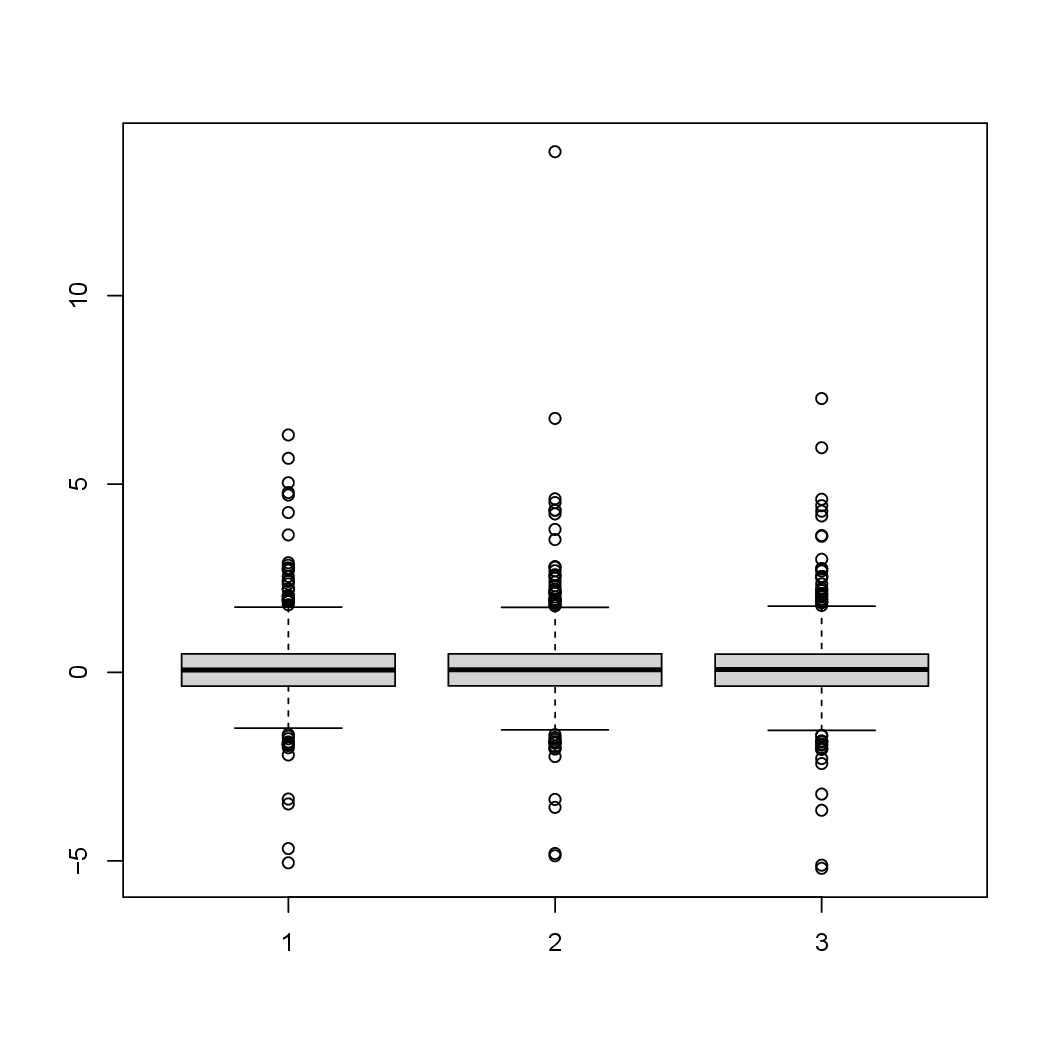}
\end{minipage}
\caption{Boxplots of the biases of the three estimators. The symbols 1, 2, and 3 in the horizontal axis denote the posterior means $\hat{\bm{\beta}}_{BR}$, $\hat{\bm{\beta}}_{BM}$, and $\hat{\bm{\beta}}_{U}$, respectively.}
\label{fig1}
\end{figure}

For example, for the posterior mean $\hat{\bm{\beta}}_{BR}$ based on the prior distribution $\pi_{BR}$, the bias is calculated by $\hat{\bm{\beta}}_{BR}-\bm{\beta}_{0}$. Similar calculations are performed for the prior distributions $\pi_{BM}$ and $\pi_{U}$. Now, we repeat the simulation 1000 times. Thus, for each true parameter and each prior, 1000 biases are computed. Figure \ref{fig1} plots the biases of the posterior means for each parameter under the three prior distributions. The left, middle, and right figures are for $\beta_1$, $\beta_2$, and $\beta_3$, respectively. Table \ref{tab1} shows the mean and standard deviation of the 1000 biases for each parameter. 

\begin{table}[htb]
\caption{Mean and standard deviation of the biases in each parameter}
\label{tab1}
\begin{center}
\begin{tabular}{lccccccccc} \hline
 & \multicolumn{3}{c}{Proposed} & \multicolumn{3}{c}{Jeffreys} & \multicolumn{3}{c}{Uniform prior} \\\cline{2-10}
 & $\beta_1$ & $\beta_2$ & $\beta_3$ & $\beta_1$ & $\beta_2$ & $\beta_3$ & $\beta_1$ & $\beta_2$ & $\beta_3$ \\\hline
Mean & \multicolumn{1}{r}{-0.878} & \multicolumn{1}{r}{0.507} & \multicolumn{1}{r}{0.101} & \multicolumn{1}{r}{-0.925} & \multicolumn{1}{r}{0.529} & \multicolumn{1}{r}{0.112} & \multicolumn{1}{r}{-0.910} & \multicolumn{1}{r}{0.526} & \multicolumn{1}{r}{0.102} \\
Stand dev & \multicolumn{1}{r}{2.195} & \multicolumn{1}{r}{1.434} & \multicolumn{1}{r}{0.862} & \multicolumn{1}{r}{2.529} & \multicolumn{1}{r}{1.521} & \multicolumn{1}{r}{0.959} & \multicolumn{1}{r}{2.212} & \multicolumn{1}{r}{1.433} & \multicolumn{1}{r}{0.880} \\\hline
\end{tabular}
\end{center}
\end{table}

From the first row of Table \ref{tab1}, we observe that the average biases of the posterior means under the proposed prior $\pi_{BR}$ take values closer to zero than the other two for all parameters. This indicates that we can obtain posterior means with less bias under the proposed prior distribution. The second row of Table \ref{tab1} gives the standard deviations of the biases, which are the standard deviations of the posterior means. 
This indicates that it is superior to the posterior mean under the Jeffreys prior $\pi_{BM}$ for the standard deviation, but when compared to the posterior mean under the uniform prior, its superiority depends on the true value. Overall,  we confirm that our proposed prior distribution gives good performance. Note that the degree of improvement of the bias becomes smaller as the sample size $n$ increases. 

\subsection{Gumbel distribution}
We consider the case when $Y_1,\cdots ,Y_n$ are i.i.d. with Gumbel distribution
\begin{align}
f(y|\mu ,\sigma )=\frac{1}{\sigma}\exp\left( -\frac{y-\mu}{\sigma}\right) \exp\left\{ -\exp \left( -\frac{y-\mu}{\sigma}\right)\right\}\qquad (-\infty <y<\infty ), \label{Gumbel_pdf}
\end{align}
where $-\infty <\mu <\infty$ and $\sigma >0$. 
It is known that the moment generating function is $M_{Y}(t):=E[\exp(tY)]=\exp (\mu t)\Gamma (1-\sigma t )$ $(t<1/\sigma)$ where $\Gamma (\cdot )$ is the Gamma function.
We assume that $\mu$ is an unknown parameter and $\sigma$ is known. 
Without loss of generality, we let $\sigma =1$. 
This density \eqref{Gumbel_pdf} is not symmetric about $\mu$. When it is symmetric, it is known that the posterior mean and the posterior mode induce unbiased estimators under the existing conditions on them. Corollary \ref{coro3} shows that the bias reduction prior $\pi_{BR}(\mu )$ is the uniform prior for $\mu$. \cite{GL11} gave $\pi_{MM}(\mu )\propto \exp(\mu/2)$ and claimed that $\hat{\mu}_{MM}-\hat{\mu}_{ML}= O(n^{-3/2})$. It follows from Proposition \ref{prop1} that $\pi_{BM}(\mu ) \propto \exp(-\mu /2)$. It is our understanding that the uniform prior is appealing in the location parameter model. The equality in Proposition \ref{prop1} holds, and this fact implies that $\hat{\mu}_{BR}-\hat{\mu}_{BM} = O_{p}(n^{-3/2})$. 

\section{Concluding Remarks}\label{sec:concluding}\label{sec6}
In this paper, we have proposed a prior distribution that removes the first-order  asymptotic bias of the posterior mean and shown that it can be derived relatively easily in several popular models. 
We conclude the paper by giving remarks on the following two points.
 In the present paper, we assumed independence for the sequence of observed random variables, but this assumption can be extended to the case where there are dependencies among random variables, as in the case of time series models. 
In addition, we have imposed some conditions to give the asymptotic expansion for the bias of the posterior mean. Although these conditions are general, studying cases where sufficient conditions are easier to check would be worthwhile.




\bibliographystyle{apalike} 

%
%

\appendix

\section{The asymptotic bias of the MLE}\label{asympt_bias_MLE}
Using the cumulants \eqref{cumulants} to rewrite the asymptotic bias of order $n^{-1}$, which is given in the equation (20) of \cite{CS68}, yields
\begin{align*}
B_{k,n}=\frac{1}{2n}\sum_{s,t,u}\kappa^{k,s}\kappa^{t,u}\left( \kappa_{s,t,u}+\kappa_{s,tu}\right) .
\end{align*}
Applying the Bartlett identity $\kappa_{stu}+\kappa_{s,tu}+\kappa_{t,su}+\kappa_{u,st}+\kappa_{s,t,u}=0$ to this equation, we have Equation \eqref{bias1}.

\section{The standard-form Laplace approximation}\label{AppenA}
This section provides sufficient conditions for the asymptotic expansion of the posterior mean $E_{\textrm{post}}[\bm{\Theta}_{k}]$. 
Since we need to distinguish between the true parameters and the components of the parameter space $\Xi$, we denote the true values by $\bm{\theta}_{0}=(\theta_{01},\ldots,\theta_{0d})^{\top}$ and the components of the parameter space by $\bm{\theta}=(\theta_1,\ldots,\theta_d)^{\top}$. 
$\mathcal{B}_{\epsilon}(\bm{\theta}_{0})$ denotes the open ball of radius $\epsilon >0$ centered at $\bm{\theta}_{0}$ in $\Xi$. For simplicity of notation, we write $\partial_{j_{1}\cdots j_{d}}=\partial^{d}/\partial \theta_{j_{1}}\cdots \partial \theta_{j_{d}}$ and $D^{2}=\partial^{2}/\partial\bm{\theta}\partial \bm{\theta}$.
Suppose that an observed random vector $\bm{Y}=(Y_1,\ldots ,Y_n)^{\top}$ has a true probability distribution $P_{\bm{\theta}_{0}}$ specified by the true parameter vector $\bm{\theta}_{0}$, which has a density $p(\bm{y}|\bm{\theta}_{0})=\prod_{i=1}^{n}p_{i}(y_{i}|\bm{\theta}_{0})$. Let $\ell_{n_{0}:n}(\bm{\theta})=\sum_{i=n_{0}}^{n}\log p_{i}(y_{i}|\bm{\theta})$ be a log-likelihood function based on partial observations $\bm{y}_{n_{0}}=(y_{n_{0}},\ldots ,y_{n} )^{\top}$. 
We consider a slightly modified version of the conditions given in \citet[pp.483-484]{KTK90}.
\begin{list}{}{}
\item{[A1]} For any $\bm{y}$ and $\bm{\theta}$, $p(\bm{y}|\bm{\theta})>0$ and for all $\bm{y}$, the log-likelihood $\ell(\bm{\theta})$ is six times continuously differentiable, and the prior $\pi(\bm{\theta})$ is four times continuously differentiable.
\item{[A2]} For all $\bm{\theta}_{0}\in \Xi$, there exist constants $\epsilon >0$ and $0<M<\infty$ such that $\mathcal{B}_{\epsilon}(\bm{\theta}_{0})\subseteq \Xi$ and for all $1\leq j_{1},\ldots ,j_{d}\leq m$ with $0\leq d \leq 6$, 
\begin{align*}
\limsup_{n\to\infty}\sup_{\bm{\theta}\in \mathcal{B}_{\epsilon}(\bm{\theta}_{0})}\left\{ \frac{1}{n}\|\partial_{j_{1}\cdots j_{d}}\ell (\bm{\theta})\| \right\} <M
\end{align*}
with $P_{\bm{\theta}_{0}}$-probability one.
\item{[A3]} For any $\bm{\theta}_{0}\in \Xi$, there exists a constant $\epsilon >0$ such that
\begin{align*}
\liminf_{n\to \infty}\inf_{\bm{\theta}\in \mathcal{B}_{\epsilon}(\bm{\theta}_{0})}\left\{ \left| \frac{-1}{n}\frac{\partial^{2}}{\partial \bm{\theta}\partial \bm{\theta}^{\top}}\ell(\bm{\theta}) \right|\right\} >0
\end{align*}
with $P_{\bm{\theta}_{0}}$-probability one.
\item{[A4]} For any $\bm{\theta}_{0}\in \Xi$ and any small $\delta >0$, there exists a nonnegative integer $n_{0}$ such that
\begin{align}
\limsup_{n\to\infty}\sup_{\bm{\theta}\in \Xi -\mathcal{B}_{\delta}(\bm{\theta}_{0})}\left\{ \frac{1}{n}(\ell_{n_{0}:n}(\bm{\theta})-\ell_{n_{0}:n}(\bm{\theta}_{0}))\right\} <0, \label{condA4}
\end{align}
with $P_{\bm{\theta}_{0}}$-probability one, and $\int \|\bm{\theta}\| \exp\{\ell_{1:n_{0}-1}(\bm{\theta})\}\pi (\bm{\theta})d\bm{\theta}$ and $\int \exp\{$\\ $\ell_{1:n_{0}-1}(\bm{\theta})\}\pi (\bm{\theta})d\bm{\theta}$ are finite with $P_{\bm{\theta}_{0}}$-probability one, where $\ell_{1:n_{0}-1}(\bm{\theta})=\sum_{i=1}^{n_{0}-1}\log p_{i}(y_{i}|\bm{\theta})$ for $n_{0}\in \{ 2,\ldots ,n\}$, and $\ell_{1:n_{0}-1}(\bm{\theta})=0$ for $n_{0}=1$ for convenience.

\end{list}
Condition [A4] ensures that the MLE $\hat{\bm{\theta}}$ is strongly consistent. If the prior $\pi (\bm{\theta})$ has a finite moment, i.e. $ \int \|\bm{\theta}\|\pi (\bm{\theta})d\bm{\theta} <\infty$, we set $n_{0}=1$. 
Equation \eqref{condA4} with $n_{0}=1$ corresponds to the consistency condition of \cite{Wa49} for the MLE. Even if the prior distribution does not have a finite moment, Condition [A4] can be satisfied by choosing an appropriate $n_0$.

The following lemma presents a valid asymptotic expansion for the posterior mean \eqref{posterior_mean}. 

\begin{lemma}\label{valid_expansion}
Under the conditions [A1]--[A4], it follows that
\begin{align*}
E_{\textrm{post}}\left[ \Theta_{k}\right] &=\hat{\theta}_{k}+\frac{1}{n}\sum_{j}\hat{h}^{kj}\left\{ \frac{\hat{\pi}_{j}}{\hat{\pi}} -\frac{1}{2}\sum_{r,s}\hat{h}^{rs}\hat{h}_{rsj}\right\} +\frac{R_{1n}}{n^{2}} , 
\end{align*}
where $R_{1n}=O_{p}(1)$.
\end{lemma}
{\sc Proof.} Because the MLE $\hat{\bm{\theta}}$ has strong consistency for the true parameter $\bm{\theta}_{0}$ and Condition [A4] with $\bm{\theta}_{0}$ replaced by $\hat{\bm{\theta}}$ holds. 
It follows from this result that, for any $\delta >0$,
\begin{align}
&\| \int_{\| \bm{\theta}-\hat{\bm{\theta}}\| >\delta} \theta_{k}\exp\{ \ell(\bm{\theta})-\ell(\hat{\bm{\theta}})\} \pi(\bm{\theta})d\bm{\theta}\| \notag \\
\leq &\exp\{-\ell_{1:n_{0}-1}(\hat{\bm{\theta}})\} \int_{\| \bm{\theta}-\hat{\bm{\theta}}\| >\delta} \exp\{ \ell_{n_{0}:n}(\bm{\theta})-\ell_{n_{0}:n}(\hat{\bm{\theta}})\} \|\theta_{k}\|\exp\{\ell_{1:n_{0}-1}(\bm{\theta})\}\pi(\bm{\theta})d\bm{\theta} \notag \\
\leq& \exp\{-\ell_{1:n_{0}-1}(\hat{\bm{\theta}})\}\exp(-nc_{1}) \int \| \theta_{k} \| \exp\{\ell_{1:n_{0}-1}(\bm{\theta})\}\pi(\bm{\theta})d\bm{\theta}\notag \\
=&O_{p}(\exp(-nc_{1})) \label{lapthm4}
\end{align}
for some $c_{1}>0$. Equation \eqref{lapthm4} corresponds to the inequality given in the proof of Theorem 4 of \cite{KTK90}. By the same argument as above, we have the equation (2.3) in Lemma 2 of \cite{KTK90}. Therefore, the lemma is proved by Theorem 4 of \cite{KTK90} . \hfill $\square$

\section{The derivation of the asymptotic bias for the posterior mean}\label{AppenB}


To obtain the asymptotic bias for the posterior mean, we add the following conditions:
\begin{list}{}{}
\item{[A5]} Observed random variables $Y_1,\ldots ,Y_n$ are independent, and the cumulants defined in \eqref{cumulants} are well-defined and have finite values for any parameter. For the cumulants with at most three subscripts, the Bartlett identities hold.
\item{[A6]} $E_{\bm{\theta}_{0}}\{ R_{1n}\} =O(1)$ as $n\to \infty$. 
\item{[A7]} $E_{\bm{\theta}_{0}}\{ R_{2n}\} =o(1)$ as $n\to \infty$ where $R_{2n}$ is defined in Equation \eqref{R2n}.
\item{[A8]} There exists a function $R_{3n}\equiv R_{3n}(\bm{\theta})$ of $\bm{\theta}$ such that for any $k=1,\ldots ,d$,
\begin{align*}
E_{\bm{\theta}_{0}}\left( \hat{\theta}_{k}-\theta_{0k}\right) =B_{k,n}+\frac{R_{3n}(\bm{\theta}_{0})}{n}
\end{align*}
and $R_{3n}(\bm{\theta}_{0}) \to 0$ as $n\to\infty$ where $B_{k,n}$ is defined in Equation \eqref{bias1}.
\end{list}
[A6] imposes a moment condition on the asymptotic error of the Laplace approximation to the posterior mean of $\Theta_k$. 
Condition [A7] imposes that the expected value of the expression \eqref{R2n} converges to zero as the sample size $n$ increases. This can be shown under condition [A5] and some moment conditions, which is not difficult but is omitted here to avoid lengthening the paper. Condition [A8] ensures that the term of order $n^{-1}$ in the bias of the maximum likelihood estimator is given by $B_{k,n}$ in Equation \eqref{bias1}, corresponding to the equation (20) of \cite{CS68}. 

{\sc Proof of Proposition }\ref{asympt.expans}.

From Lemma \ref{valid_expansion}, we obtain a valid asymptotic expansion for the posterior mean $E_{\textrm{post}}[\Theta_k]$.
From this, we can show Theorem \ref{asympt.expans} by following the derivation of the asymptotic bias in Section \ref{sec2}. \hfill $\square$

\section{Proofs for the Corollaries}\label{proofs}

{\sc Proof of Corollary} \ref{coro1}.

As $Y_1,\ldots ,Y_n$ are i.i.d., we have
\begin{align*}
\kappa_{1,1}&=\frac{1}{n}E_{\theta}\left\{ \left( \frac{d}{d\theta}\ell (\theta )\right)^2\right\}=-E_{\theta}\left\{ \frac{d^{2}}{d\theta^2}\log p(Y_{1}|\theta )\right\} =I_{1}(\theta),\notag \\
\kappa_{111}&=E_{\theta}\left\{ \frac{d^{3}}{d\theta^3}\log p(Y_{1}|\theta )\right\} ,
\intertext{and}
\kappa_{1,11}&=E_{\theta}\left\{ \frac{d}{d\theta}\log p(Y_{1}|\theta )\frac{d^{2}}{d\theta^2}\log p(Y_{1}|\theta )\right\} .
\end{align*}
Hence, the second term on the right-hand side of Equation \eqref{main_result} becomes
\begin{align}
\sum_{r,s}\kappa^{r,s}(\kappa_{rsj}+\kappa_{r,js})&=\frac{1}{I_{1}(\theta )}\left\{ E_{\theta}\left( \frac{d^3}{d \theta^3}\log p(Y_{1}|\theta) \right) +E_{\theta}\left( \frac{d}{d \theta}\log p(Y_1|\theta )\frac{d^2}{d \theta^2}\log p(Y_1|\theta ) \right)\right\} . \label{coro1_proof}
\end{align}

Using the condition on the interchange of integral and derivative, we have
\begin{align*}
\frac{d}{d \theta}E_{\theta}\left( \frac{d^2}{d \theta^2}\log p(Y_{1}|\theta) \right) =E_{\theta}\left( \frac{d^3}{d \theta^3}\log p(Y_{1}|\theta) \right) +E_{\theta}\left( \frac{d^2}{d \theta^2}\log p(Y_{1}|\theta)\frac{d}{d \theta}\log p(Y_{1}|\theta) \right) .
\end{align*}
Applying this result to Equation \eqref{coro1_proof} yields 
\begin{align*}
\frac{d}{d \theta}\log\pi (\theta )=\frac{d}{d \theta}\log I_{1} (\theta ),
\end{align*}
which completes the proof. \hfill $\square$

\hspace*{-1.5em}{\sc Proof of Corollary }\ref{coro2}

To clarify the dependence of $\kappa_{r,s}$ and $\kappa_{rsj}$ on parameter $\bm{\theta}$, we rewrite $\kappa_{r,s}=\kappa_{r,s}(\bm{\theta})$ and $\kappa_{rsj}=\kappa_{rsj}(\bm{\theta})$. 
The symbol $\mathrm{tr}(\bm{A})$ denotes the trace of matrix $\bm{A}$.

Because $(\partial /\partial \theta_{j})\kappa_{rs}(\bm{\theta})=\kappa_{rsj}(\bm{\theta})$ from assumption (C), we have
\begin{align}
\sum_{r,s}\kappa^{r,s}\kappa_{rsj}&=\sum_{r,s}\kappa^{r,s}(\bm{\theta})\left( \frac{\partial}{\partial \theta_{j}}\kappa_{rs}(\bm{\theta})\right) \notag \\
 &=-\sum_{r,s}\bm{I}^{r,s}(\bm{\theta})\left( \frac{\partial}{\partial \theta_{j}}\bm{I}_{rs}(\bm{\theta})\right) \notag \\
 &=-\textrm{tr}\left\{ \bm{I}(\bm{\theta})^{-1}\frac{\partial}{\partial \theta_j}\bm{I}(\bm{\theta})\right\} \notag \\
 &=-\frac{\partial}{\partial \theta_j}\log | \bm{I}(\bm{\theta} )| ,\notag 
\end{align}
which completes the proof. \hfill $\square$

\hspace*{-1.5em}{\sc Proof of Corollary \ref{coro3}}. The assumption yields that $(\partial /\partial \theta_r )\kappa_{sj}=0$, which is rewritten as 
\begin{align*}
0&=\frac{\partial}{\partial \theta_r }E_{\bm{\theta}}\left[ \frac{\partial}{\partial \theta_{s}\partial \theta_{j}}\ell (\bm{\theta})\right] \\
&=\int \partial_{rsj}\ell(\bm{\theta}) p(\bm{y}|\bm{\theta})d\bm{y} +\int \partial_{sj}\ell(\bm{\theta})\partial_{r} \ell(\bm{\theta})p(\bm{y}|\bm{\theta})d\bm{y}.
\end{align*}
This implies that $\kappa_{rsj}+\kappa_{r,js}=0$, which indicates that the second term in the right-hand side of Equation \eqref{def_eq1} vanishes. \hfill $\square$

\end{document}